\documentclass[conference]{IEEEtran}
\IEEEoverridecommandlockouts
\usepackage{cite}
\usepackage{amsmath,amssymb,amsfonts}
\usepackage{algorithmic}
\usepackage{graphicx}
\usepackage{textcomp}
\usepackage{xcolor}
\usepackage{physics}

\def\BibTeX{{\rm B\kern-.05em{\sc i\kern-.025em b}\kern-.08em
    T\kern-.1667em\lower.7ex\hbox{E}\kern-.125emX}}
\begin{document}
\newcommand{\RR}{\mathbb{R}}
\newcommand{\CC}{\mathbb{C}}
\newcommand{\dsp}{\displaystyle}

\title{SurfaceNet: Fault-Tolerant Quantum Networks with Surface Codes
}
\author{Tianjie Hu, Jindi Wu, and Qun Li
\thanks{* Tianjie Hu, Jindi Wu, and Qun Li are with the Computer Science Department at William \& Mary.}}

\maketitle

\thispagestyle{plain}
\pagestyle{plain}




\begin{abstract}
Quantum networks serve as the means to transmit information, encoded in quantum bits or qubits, between quantum processors that are physically separated. Given the instability of qubits, the design of such networks is challenging, necessitating a careful balance between reliability and efficiency. Typically, quantum networks fall into two categories: 
those utilize quantum entanglements for quantum teleportation, and those directly transfer quantum message.
In this paper, we present SurfaceNet, a quantum network in the second category that employs surface codes as logical qubits for preserving and transferring message. Our approach of using surface codes can fault-tolerantly correct both operational and photon loss errors within the network. We propose a novel one-way quantum communication procedure, designed to better integrate surface codes into our network architecture. We also propose an efficient routing protocol that optimizes resource utilization for our communication procedure. Simulation results demonstrate that SurfaceNet significantly enhances the overall communication fidelity.
\\
\end{abstract}


\section{Introduction}

Quantum networks enable the transmission of quantum information across extended distances, facilitating the collaborative use of quantum computing resources in diverse locations. These networks find applications in various domains such as distributed quantum machine learning, quantum cryptography, quantum key distribution, and more. However, given the nature of quantum mechanics, the design of quantum networks is nontrivial, with the challenge of balancing reliability and efficiency being a significant consideration. Over the past decades, many research groups have focused on this field, proposing various quantum network structures that can generally be categorized into two schemes based on their utilization of quantum entanglement.

The first scheme~\cite{briegel1998quantum, chen2022heuristic, zeng2022multi, li2022fidelity} of quantum networks utilizes quantum entanglements for quantum teleportation between nodes. As quantum data are highly unstable and sensitive to the environment, in this scheme, a series of quantum entangling and swapping is performed along the path before the actual data are transferred. Once all the swapping is completed, the sender and receiver would attain a pair of entangled qubits, which can be used for quantum teleportation to transfer the actual data. However in practice, this scheme is often very inefficient, as the entanglement generation is a probabilistic process and the lifespan of the generated pairs is considerably short.
\textcolor{black}{The method of entanglement purification has been proposed to improve the reliability of the network. It can combine several low-fidelity entanglements into one with higher fidelity, but in return it adds more burden to the computing resources.}
In addition, a large amount of classical communications across nodes are needed for both quantum teleportation and entanglement purification, which often dominates the total time of a single communication. 

The second scheme~\cite{fowler2010surface,muralidharan2014ultrafast} of quantum networks transfers quantum message directly between nodes, and often utilizes logical qubits to carry and preserve the message data.
In this scheme, the actual messages are transferred directly as in the classical network. 
\textcolor{black}{To address the instability of qubits, a single logical qubit is encoded or represented by multiple physical qubits connected by quantum gates, and the state of the logical qubit is determined collectively by these physical qubits. 
For example, logical qubits can be constructed by a geometric layout of quantum entanglements, replacing the unstable physical qubits. } 
This scheme provides resilience against random noise via the intrinsic quantum gates and redundant physical qubits within a logical qubit. 
However, it comes with the drawback of significantly increasing the network's traffic since to transmit a single logical qubit, we need to transmit multiple physical qubits. 

\begin{figure}[t]
\color{black}
\centerline{\includegraphics[scale=.33]{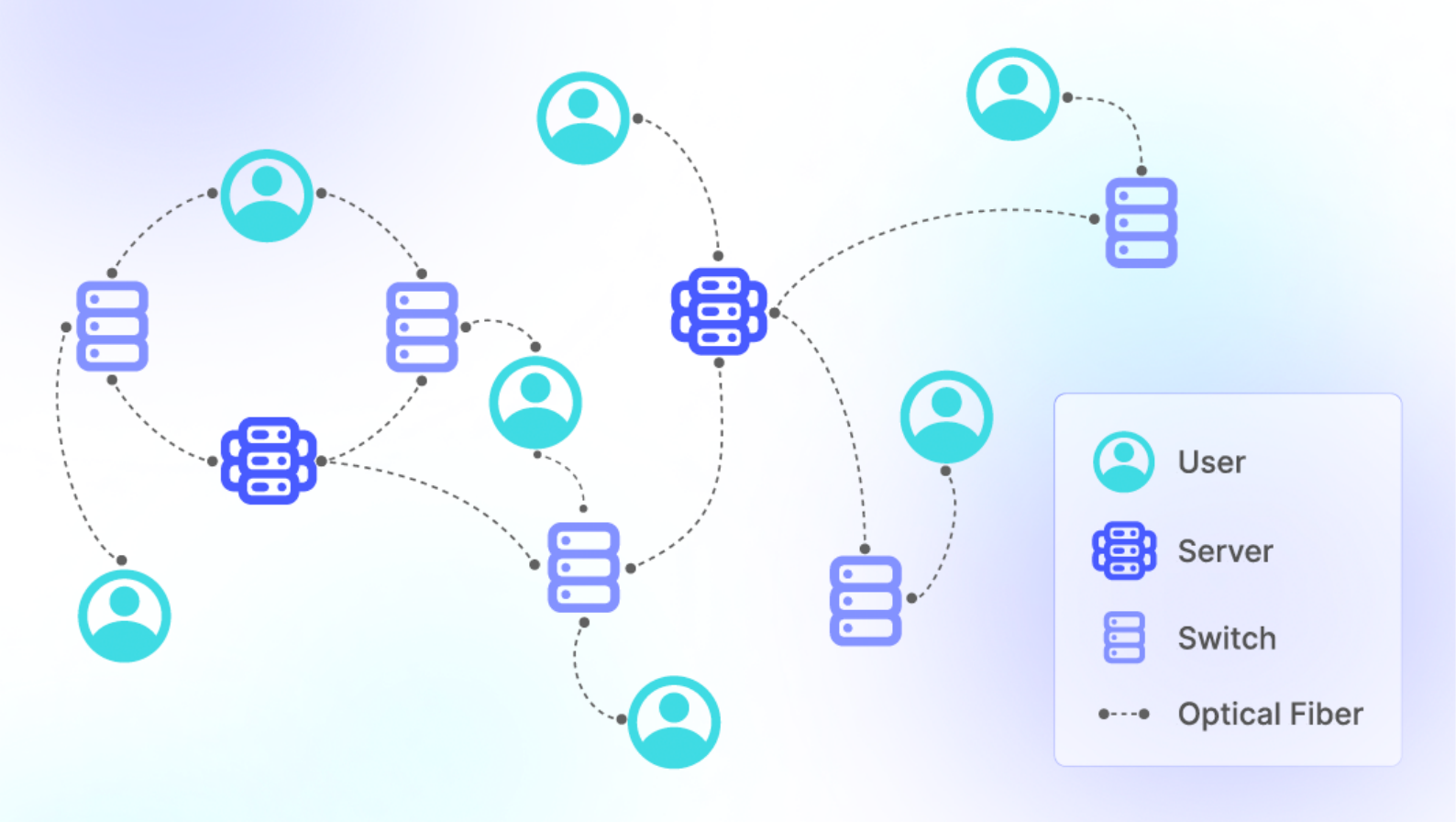}}
\caption{Example quantum network of SurfaceNet, where users communicate with each other through switches and servers, interconnected by optical fibers.}
\label{fig:network}
\end{figure}

In this paper, we propose a quantum networking strategy which fits in the second scheme of quantum networks, utilizing surface code as the logical qubits. There are two primary motivations behind this design choice. 
Firstly, surface code shows promise as a future quantum network technique with its compact size and high error threshold. Extensive research~\cite{google2023suppressing} has been conducted on the design and application of surface codes. 
Secondly, integrating surface codes with the second scheme of quantum networks offers several advantages over existing approaches. One notable advantage is the elimination of classical communication, which typically occupies a significant portion of the total communication time. 
Furthermore, as detailed in this paper, it provides insights for a new structure of quantum networks: 
\textcolor{black}{for enhancing the network efficiency, individual surface codes can be partitioned into smaller parts and transmitted separately; and for enhancing the network reliability, the intrinsic entanglements of surface codes can be leveraged to perform error corrections at designated intermediate nodes.} 
These findings lay the foundation for potential future research directions in the area of quantum networks.

Our main contributions are summarized as follows:

\begin{itemize}
\item We present the first quantum network with surface codes, which directly employs surface codes for communication and eliminates the need for inter-node entanglements. 
\item We propose an efficient surface code based communication procedure specifically designed for our network. This process involves partitioning the surface codes and transferring each part along separate paths.
\item We formulate the routing protocol of our network as an integer programming problem, and solve it in its linear relaxation form. We evaluate the performance of our network against our proposed baseline models and the popular entanglement purification networks.
\\
\end{itemize}

\section{Background}

\subsection{Qubits}

Different from bits (either 0 or 1) used for classical computing, qubits used for quantum computing exist as linear combinations of the two basis states $\ket{0}$ and $\ket{1}$. Known as its property of superposition, qubit lives in a continuum of these two basis states, and it is only upon a quantum measurement that we can determine its sole value of being either 0 or 1. 
Via this property, we can use each qubit to simultaneously carry and function as both values (0 and 1) of classical information. Furthermore, when we use a group of $n$ qubits as input, this advantage accumulates exponentially because they can collectively represent $2^n$ classical information states.
However, this nature of qubit also makes it highly unstable, suffering from noises from the environment and random collapsing. Thus, instead of directly using the unstable physical qubits, many papers have proposed designs of quantum codes, which are composed of clusters of physical qubits 
\textcolor{black}{and function as single logical qubits~\cite{gottesman1997stabilizer}. Serving as a logical qubit, each quantum code consists of two parts: data qubits for storing information, and ancillary qubits for maintaining information integrity. Both parts are composed of physical qubits interconnected with quantum gates.}

\subsection{Surface Codes}

\begin{figure}[t]
\centerline{\includegraphics[scale=.48]{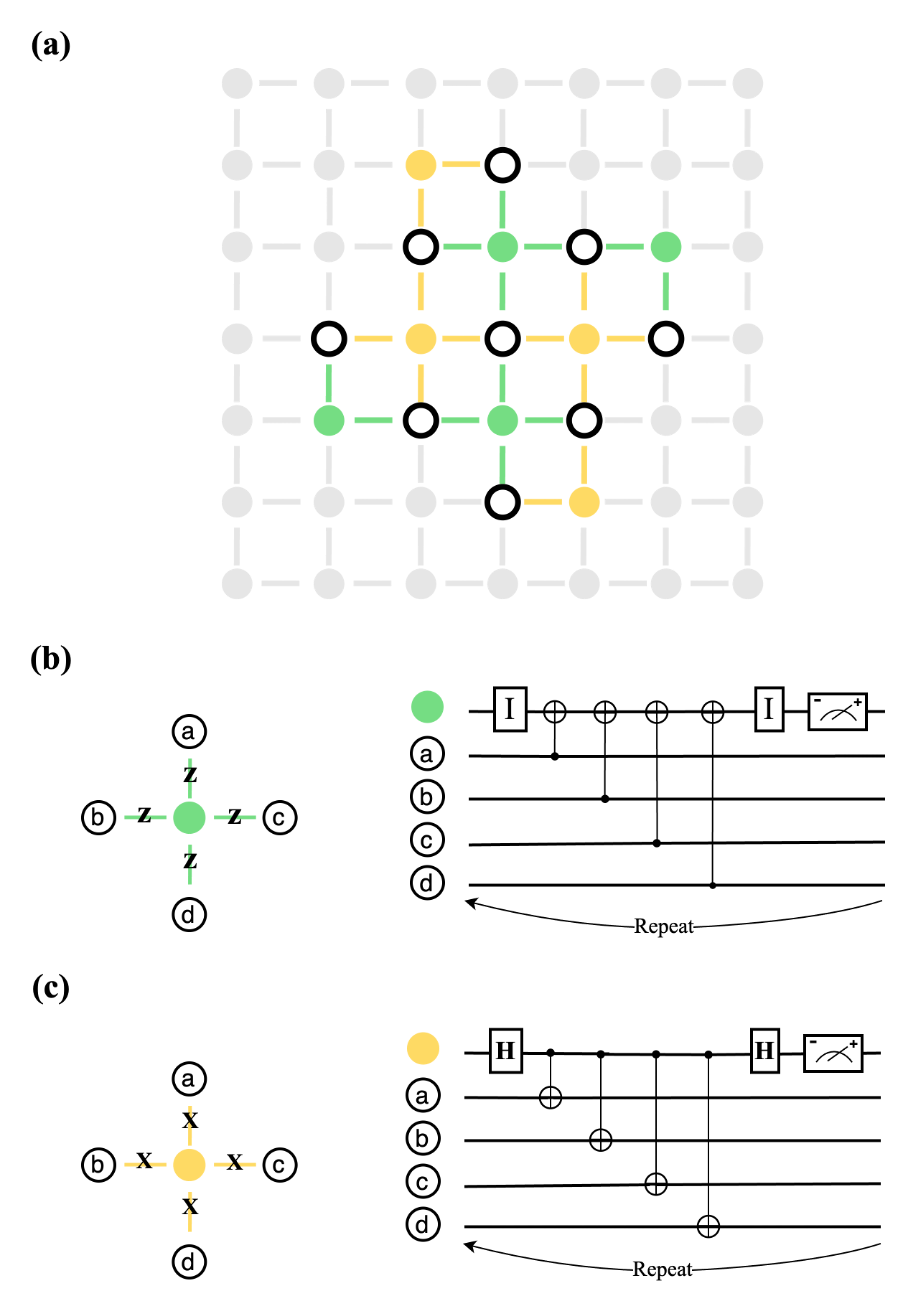}}
\color{black}
\caption{ (a) Example distance-3 surface code. Open circles represent data qubits, and filled circles represent measurement qubits (green for Measure-Z, yellow for Measure-X).\quad (b) Quantum circuit on Measure-Z for extracting X and Y types errors.\quad (c) Quantum circuit on Measure-X for extracting Y and Z types errors.}
\label{fig:sfcode}
\end{figure}

Among the different kinds of quantum codes, surface code has been the most promising one, for its compact size and high error threshold. 
As in Fig.~\ref{fig:sfcode}(a), surface codes are composed of two groups of physical qubits: data qubits and measurement qubits.
Fundamentally, data qubits and measurement qubits are of no difference -- they are all physical qubits -- yet they differ significantly in their roles within surface code: the logical qubit information is encoded and stored within data qubits, while measurement qubits serve as ``monitors" of this information in each error correction cycle. Each data qubit is connected to its neighboring measurement qubits: two measure-X qubits, and two measure-Z qubits, except on boundaries. If an error occurs in a specific data qubit, its neighboring measurement qubits would reflect the error: the measure-X qubits would reflect any Y or Z errors, and the measure-Z qubits would reflect any X or Y errors. Thus, measurement qubits are also referred to as syndrome qubits, or in the code theory, as ancilla qubits.

The mechanism behind surface codes is based on periodic quantum entanglements and reduction on degrees of freedom, and we refer readers to~\cite{nielsen2002quantum} for a great introduction to Quantum Computation and Quantum Information, and to~\cite{fowler2012surface} for a detailed explanation on Surface Codes.

\subsection{Error Correction}

Surface codes are endowed with the capability for error correction during their initialization process. To initialize a surface code, we firstly prepare it with only the data qubits: without loss of generality, we commonly prepare all data qubits to be in $\ket{0}$. Next, we complete the surface code by putting on the measurement qubits and connect them to their neighboring data qubits with quantum CNOT gates (as in Fig.~\ref{fig:sfcode}(b),~\ref{fig:sfcode}(c)): again without loss of generality, we commonly prepare all measurement qubits also in their $\ket{0}$ states. After all CNOT gates are established, the states of all measurement qubits would change accordingly: as we prepare all qubits in $\ket{0}$, the initial measurement results of all measure-Z qubits should retain in their $\ket{0}$; meanwhile, as each qubit is generated with random phases, the initial measurement results of the measure-X qubits would be random. 
Yet, we can still force all measure-X results to be in $\ket{-}$ by applying rotation-X gates accordingly to the data qubits, without twisting their encoded logical information~\cite{wootton2017proposal}. The advantage of this approach is that there is no longer a need to store the measurement results for each round. Once all is done, a surface code is effectively initialized. 
If no error occurs in the future, all subsequent measurements on the measurement qubits would hold the same: $\ket{0}$ for all measure-Z, and $\ket{-}$ for all measure-X. And if the measurement result of any measurement qubit changes, then by the mechanism of surface code, at least one of its neighboring data qubits has error.

\begin{figure}[t]
\color{black}
\centerline{\includegraphics[scale=.5]{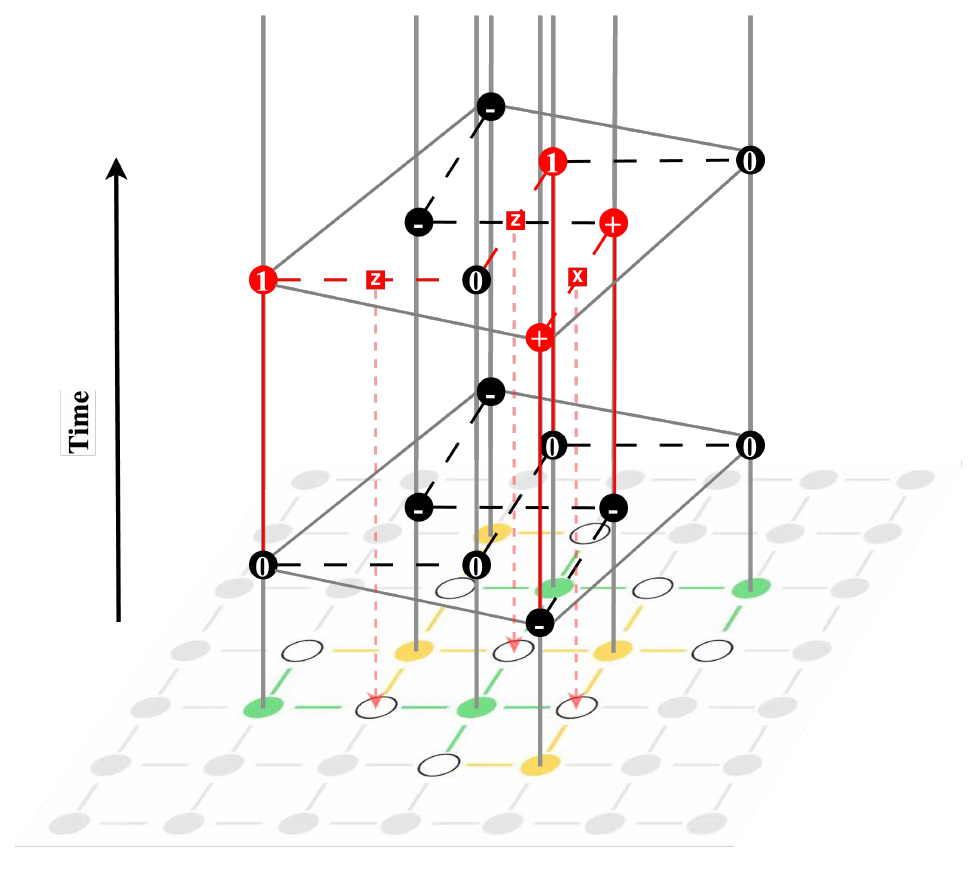}}
\caption{Evolution of surface code error correction cycles. Error maps at two consecutive timestamps are shown, in which the first one (lower layer) exhibits no error and the second one (higher layer) exhibits an error map of four syndromes. In the higher layer, the four flipped measurement qubits are marked red, and the red dashed-lines form a possible inference of errors, where the red dashed-arrows point out the three data qubits with errors.}
\label{fig:decoder}
\end{figure}

\begin{figure*}[t]
\color{black}
\centering
\includegraphics[scale=.56]{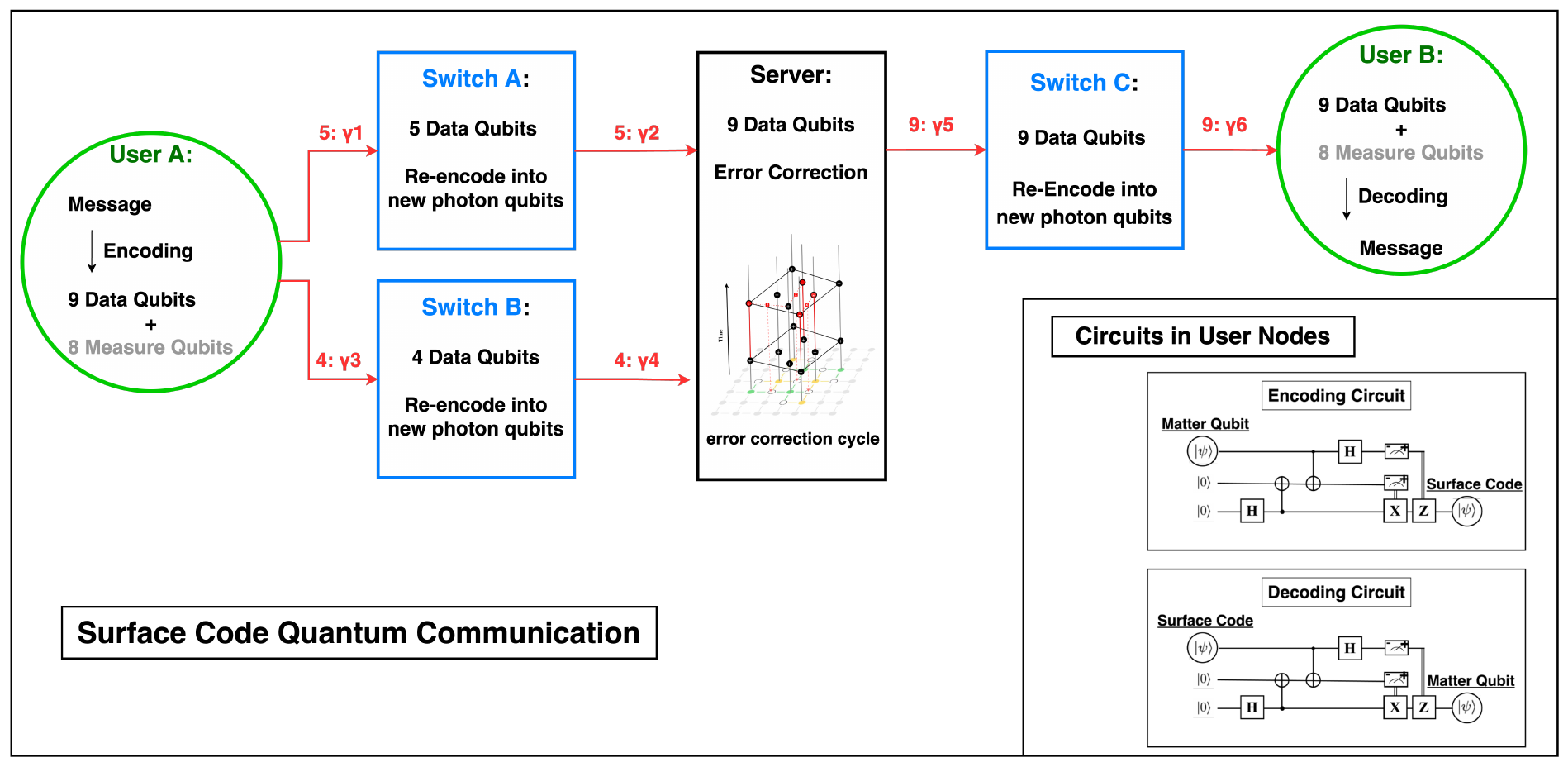}
\caption{Example one-way communication of transferring 9 data qubits from User A to B, where the green circles represent users, blue squares represent switches, black squares represent servers, and red arrows represent optical fibers. Within each node, the corresponding jobs are shown in the figure, and each optical fiber is labeled with its own fidelity and the corresponding number of transpassing qubits. Along the routing path, only data qubits are transferred.}
\label{fig:oneway}
\end{figure*}

After the initialization, repetitive error correction cycles are performed to maintain the integrity of surface codes, via periodically detecting and offsetting any errors before they accumulate. Within each cycle, a two-step process is performed. First, all measurement qubits are measured, 
\textcolor{black}{via projecting them~\cite{fowler2012surface} into one of the two eigenstates of pauli-Z or pauli-X matrices: $\ket{0}$ and $\ket{1}$ for pauli-Z, $\ket{-}$ and $\ket{+}$ for pauli-X. As discussed above, normally we initialize all measure-Z qubits to be in $\ket{0}$ and all measure-X qubits to be in $\ket{-}$. Thus, if we see a measurement result of $\ket{1}$ on a measure-Z qubit or a $\ket{+}$ on a measure-X qubit, we define it as a ``syndrome". Within the same error correction cycle, all syndromes collectively form an ``error map" (as in Fig.~\ref{fig:decoder}) of all data qubits.} 
Next, an error correction decoder is applied to the error map to infer a possible error pattern among the data qubits. Given the nature of qubits, we can never know the real distribution of errors, since it requires directly measuring the data qubits, which would destroy their encoded logical information and collapse them down to classical bits. 
\textcolor{black}{Yet when the accumulated error is below certain thresholds, the decoders can always give a pattern of possible errors that is equivalent upon stabilizers to the real error pattern. Thus, error correction cycles need to be scheduled in a regular basis before error accumulates, and more frequently when the noise level of the environment is high. At each time of performing an error correction cycle, the decoder aims to connect all syndromes either into pairs or to the boundary, so this inference problem can be formulated as a Minimum Weight Perfect Matching (MWPM) problem.}
As MWPM has been widely studied, many well-performing algorithms and heuristic methods in this area can be directly applied to surface code error correction. In addition, other inference methods such as the Union-Find decoder, Monte Carlo method, Reinforcement Learning model, and Tensor Network techniques. have also been proposed, with competitive performances in different scenarios and error thresholds.
\\

\section{Quantum Network Model}

In this section, we present our quantum network in detail. First, we introduce the five main components in our network. Next, we illustrate how each surface code based communication is performed. Finally, we explain how our network routing operates and consolidate it to be a mathematics formulation. 

\subsection{Components}

\textbf{Users } User nodes generate communication requests in the network, which contain arbitrary number of surface codes to be transferred.

\textbf{Switches } Switch nodes serve as the intermediate stations along each communication path. Within each switch, the incoming data qubits are refreshed and encoded into new photons, to reduce the chance of random decoherence. 

\textbf{Servers } Server nodes are switch nodes that not only refresh data qubits, but also are capable of performing error correction cycles on them. 

\textbf{Center } The center uses classical communication to collect requests and status of the network. Previously we mentioned that classical communication can be completely eliminated in our network, but we still include the center node at the cost of a slight overhead, for simplifying the routing process.

\textbf{Optical Fibers } Optical fibers are edges in the network that interconnect users or switches.

\subsection{Surface Code Based Communication}

Each pair of users can communicate either directly or via intermediate switches. At the sender node of each communication, the message can be either individual qubits or surface codes. In the case of individual qubits, the message can be encoded into surface codes via the Encoding Circuit as in Fig.~\ref{fig:oneway}; and for message in the forms of surface codes, the Encoding Circuit can also be utilized to encode it into surface codes of larger sizes for better fault tolerance. And during the encoding process, all physical data qubits are also converted into photons for transmission along optical fibers. After the encoding, only the data qubits are transferred, and the measurement qubits are retained in the sender node.
Next, during the transmission along routing paths, the data qubits can be partitioned into parts and transferred separately through different switches or servers. Notably, if an error correction cycle is scheduled along a communication path, all data qubits of the same surface code need to go through and be present in the same server. 
Finally at the receiver node of each communication, the transferred data qubits, together with newly prepared measurement qubits, are decoded back to the message via the Decoding Circuit as in Fig.~\ref{fig:oneway}. 

For example, in Fig.~\ref{fig:oneway}, a surface code containing nine data qubits needs to be transmitted from User A to User B. 
As determined by the routing center, it is partitioned into two parts: one part contains five data qubits, and the other part contains four data qubits. 
During the transmission, the part with five data qubits qubits travel through the upper switch, while the rest four travel through the lower switch. Within each switch, each transpassing data qubit is only refreshed and no error correction is performed.
Next we have these two paths merge into the server in the middle, where an error correction cycle is performed.
After the error correction, all nine data qubits are transmitted together via the last switch to the receiver node. Note that the nine data qubits can be partitioned again after the error correction, but for illustrative purpose, we let them being transmitted together here.
\\

\subsection{Network Routing}

Based on this surface code communication, we aim to construct a reliable and efficient network, which can achieve both a high communication success rate and a high communication throughput. And we make the following assumptions:

\begin{itemize}
\item The network is bidirected and connected, so there exists at least one path from one user to another;
\item The fidelity of each optical fiber can be measured, and does not change during the routing;
\item All switch nodes and optical fibers have their own capacities, which are not trivially large.
\end{itemize}
\textcolor{black}{
Meanwhile, the placement of servers within the network plays a critical role in the overall performance of the network. For this paper, we assume their locations to be fixed.
}
During each round of routing, the center collects information from users and switches to schedule the routing. And we formulate the routing problem as an optimization problem, whose objective function is the number of concurrent communications which needs to be maximized to achieve the highest communication throughput. Recall the other goal of our network is to achieve a high communication success rate, so we contain a fidelity constraint within our routing formulation to ensure that each communication maintains a high success rate. 

\textcolor{black}{
The fidelity constraint requires the end-to-end fidelity of each routing path to be above a specified threshold. As in Fig.~\ref{fig:oneway}, the fidelity of each optical fiber can be measured as $\gamma_i \in [0,1]$, in which higher fidelity means lower probability of random errors. To calculate the end-to-end fidelity, we need to account for all optical fibers in the path, interconnected either by switches or servers.
At each switch, each transpassing data qubit is only refreshed and no error correction is performed, so the concatenation of two optical fibers via a switch is calculated as multiplying their individual fidelity. For example, in Fig.~\ref{fig:oneway}, each of the five qubits through the upper switch has a fidelity of $\gamma_1*\gamma_2$ when it arrives at the server, while each of the four qubits through the lower switch has a fidelity of $\gamma_3*\gamma_4$ when it arrives at the server.
When an error correction is performed at a server, we average the individual fidelity of data qubits within the same surface code plus an added $\omega$, which serves as a reward boost to the fidelity resulted from performing error correction. For example, in Fig.~\ref{fig:oneway}, the surface code after the error correction in the server has a fidelity of 
\begin{equation}
[\frac{5}{9}(\gamma_1*\gamma_2)+\frac{4}{9}(\gamma_3*\gamma_4) + \omega ]
\label{eq:fid_err}
\end{equation}
since it has five qubits through the upper switch and four qubits through the lower switch, and the $\omega$ represents the reward for performing error correction. The added number is divided by nine since in total there are nine data qubits.
}

There are other constraints in our formulation for setting up and adapting the routing process for different networks, which will be discussed extensively in the following section. 
\\

\begin{figure*}[t]
    \color{black}
\centering
\fbox{\includegraphics[scale=1.05]{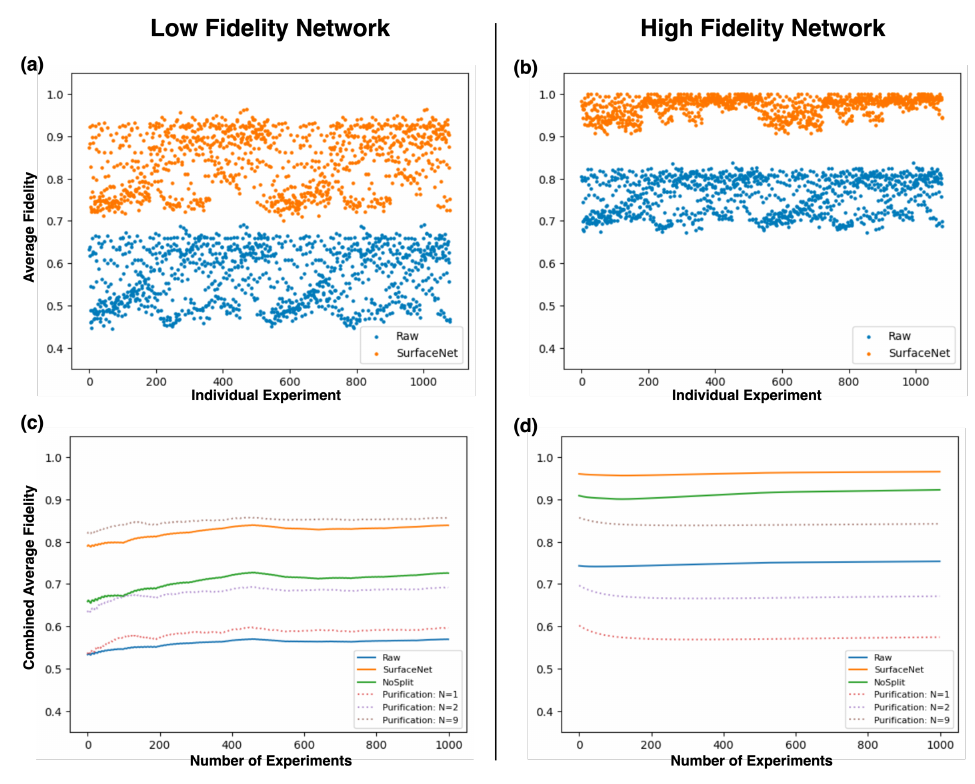}}
\caption{(a) (b) In-place comparisons of each experiment between our SurfaceNet model and the baseline model Raw. The average capacity is calculated by communication in each accounted experiment. (c) (d) Comparison of all five models. The orange line represents SurfaceNet, the blue and green lines represent the two baseline models, and the three dimmed lines (red, purple, brown) represent Entanglement Purification networks, using different numbers of entanglement pairs for each purification. The combined average fidelity is calculated by communication in all accounted experiments. }
\label{fig:eval}
\end{figure*}

\subsection*{Mathematics Formulation}

Before we proceed to the actual formulation of our routing protocol, we make the following definition as in Table~\ref{table:notation}.

\begin{table}[h]
\begin{center}
\caption{\label{table:notation}TABLE OF NOTATIONS USED IN ROUTING FORMULATION}
\begin{tabular}{|| l l ||} 
 \hline\hline
 Notation & Definition \\ 
 \hline\hline
 $\mathbb{U}$ & The set of all Users \\
 $\mathbb{R}$ & The set of all Switches (including Servers) \\
 $\mathbb{R}\mathbb{R}$ & The set of all Servers \\
 $\eta_r$ & Capacity of each switch $r\in \mathbb{R}$ \\
 $E$ & The set of all Edges \\
 $\eta_e$ & Capacity of each edge $e\in E$ \\
 $\gamma_e $ & Fidelity of each edge $e\in E$\\
 $\omega$ & Fidelity reward of performing error correction in a server \\
 $\Gamma$ & Fidelity threshold of each request \\
 $\mathbb{K}$ & The set of all communication requests $k=[(s,d),n,m_k]$\\
 $n$ & Number of data qubits in each surface code\\
 $m_k$ & Number of surface codes in Request $k$\\
 $Y_k$ & Integer variable of determining execution of request $k$\\
 $y^k_e$ & Integer variable of determining number of qubits for request \\
 & $k$ travelling through edge $e$ \\ 
 $x^k_r$ & Integer variable of determining number of error corrections\\
 & scheduled for request $k$ in each server $r$\\
 \hline
\end{tabular}
\end{center}
\end{table}

Then our formulation is stated as follows:

\begin{equation}
\label{eq:obj}
\max \sum_{k\in \mathbb{K}} Y_k
\end{equation}

subject to:
\begin{equation}
\begin{aligned}
\label{eq:ints}
&\ Y_k\in [0,m_k] \textbf{ , } x^k_r\in [0,m_k] \textbf{ , } y^k_e\in [0,n*m_k] \\[0.5ex]
\end{aligned}
\end{equation}

\begin{equation}
\begin{aligned}
\label{eq:cons}
&\ \sum_{e\in E}\gamma_e*y^k_e - \omega\sum_{r\in \mathbb{R}\mathbb{R}} x^k_r \le \Gamma *Y_k & \forall_{k\in \mathbb{K}} &\\
&\ \sum_{(s,j)\in E} y^k_{(s,j)} = n * Y_k & \forall_{k=(s,d)\textbf{ }\in \mathbb{K}} &\\
&\ \sum_{(i,s)\in E} y^k_{(i,s)} = 0 & \forall_{k=(s,d)\textbf{ }\in \mathbb{K}} &\\
&\ \sum_{(d,j)\in E} y^k_{(d,j)} = 0 & \forall_{k=(s,d)\textbf{ }\in \mathbb{K}} &\\
&\ \sum_{(i,d)\in E} y^k_{(i,d)} = n * Y_k & \forall_{k=(s,d)\textbf{ }\in \mathbb{K}} &\\
&\ \sum_{(i,r)\in E} y^k_{(i,r)} = \sum_{(r,j)\in E} y^k_{(r,j)} & \forall_{k\in \mathbb{K}} \textbf{ } \forall_{r\in\mathbb{R}} &\\
&\ \sum_{(i,r)\in E} \sum_{k\in\mathbb{K}} y^k_{(i,r)} \le \eta_r & \forall_{r\in\mathbb{R}} &\\
&\ \sum_{(i,r)\in E} y^k_{(i,r)} = n * x^k_r & \forall_{k\in \mathbb{K}} \textbf{ } \forall_{r\in\mathbb{R}\mathbb{R}} &\\
&\ \sum_{k\in\mathbb{K}} y^k_{e} \le \eta_e & \forall_{e\in E} &\\
\end{aligned}
\end{equation}
\\

\textcolor{black}{
{\bf Objective Function: } 
As in the last few rows in Table~\ref{table:notation} and Eq.~\ref{eq:ints}, the variables in our routing formulation are three sets of integer variables: $Y_k$ to decide whether each request will be scheduled or partly scheduled, $y_e^k$ to decide all the resources scheduled for each request, and $x_r^k$ to decide if error correction will be scheduled for each request needs along the path. And the Objective Function in our formulation is rather straightforward as in Eq.~\ref{eq:obj}, which aims to maximize the total communication throughput in our network. 
}
\\

\textcolor{black}{
{\bf Fidelity Constraints: } Recall the other goal of our routing is to maintain a high success rate for each communication, so we put it as the first set of constraints as in Eq.~\ref{eq:cons}. As discussed above, we calculate the fidelity of each surface code through the routing path as either concatenation or combined average of the fidelity at each path segment, and we require it to be higher than a specified fidelity threshold.
For example, the result fidelity of the transferred surface code in Fig.~\ref{fig:oneway} is:
\begin{equation}
[\frac{5}{9}(\gamma_1*\gamma_2)+\frac{4}{9}(\gamma_3*\gamma_4) + \omega ] * \gamma_5*\gamma_6 \ge \Gamma
\label{eq:fid}
\end{equation}
in which the square brackets represent the process of error correction, and the threshold $\Gamma$ on the right side can be adjusted for controlling the success rate of communication. 
}
\\

{\bf Flow Constraints:}
The next five sets of constraints in Eq.~\ref{eq:cons} follow directly from a standard network flow formulation. 
First, we set up the constraints for the Source and the Sink for a request.
At the source node of each request $k$, we restrict its sum of all outgoing flows to be $Y_k$ multiplied by the size of the transferred surface code. And we restrict its sum of all incoming flows to be $0$, to force all its sent message to reach the sink node successfully. Similarly, at the sink node of each request $k$, we restrict its sum of all incoming flows to be $Y_k$ multiplied by the size of the transferred surface code, and its sum of all outgoing flows to be $0$. 
Next, we set up the node Conservation constraint at each node except the source and sink nodes, which restricts its sum of all incoming flows to be equivalent to its sum of outgoing flows.
\\

{\bf Capacity Constraints: }
The last three sets of constraints in Eq.~\ref{eq:cons} are capacity constraints for our network, correspondingly for switches, servers, and optical fibers.
At each switch, we sum all incoming flows of all requests and restrict the summation to be smaller than the current capacity at the switch. 
Next, at each server, despite its capacity constraint as a normal switch, we also need to apply a constraint for the error correction performed on it. For each request $k$, we restrict the sum of all incoming flows to this server $r$ to be the size of the transferred surface code multiplied by $x_r^k$.
Finally, at each optical fiber, we sum the transpassing flows of all requests and restrict the summation to be smaller than the capacity of the optical fiber. 
\\

In total, we have three sets of integer variables and nine sets of constraints for our routing formulation. As Integer Programming problems are NP-hard problems and do not guarantee an optimal solution in polynomial time, 
we use its Linear Programming Relaxation form with rounding in the Evaluation section to evaluate its performance against the baseline models.
\\



\section{Evaluation}

For comparison, we use our routing formulation of ``SurfaceNet" as in the previous section, but relax its integer constraints to transform it to an linear programming formulation, and we apply rounding to its solutions. We compare its performance against two proposed baseline models: the first one is ``Raw", which does not perform error corrections along the routing path; and the other is ``NoSplit", which can only transfer each surface code completely along single paths. We also include a comparison against the popular entanglement purification networks~\cite{li2022fidelity}, in which the entanglement purification technique is applied at each edge of the network to increase its corresponding fidelity. We choose three entanglement purification networks, namely ``N$=$1", ``N$=$2", ``N$=$9", in which N indicates the number of entanglement pairs required for each time of entanglement purification at each edge.

For simulation, we use the networkx library in Python to randomly generate Barabasi-Albert graphs as our small-world networks of 20 nodes. We assign the four most connected nodes to be server nodes, the ten least connected nodes to be user nodes, and the rest to be switch nodes. The fidelity assignments to edges are random but adjusted for calculation. For a realistic simulation of the real world setting, we choose the range of fidelity to be $[0.75,1]$ and $[0.5,1]$, correspondingly for the High Fidelity Network and the Low Fidelity Network. Evaluation results show similar trends for slightly broader or narrower ranges, yet extreme scenarios such as $[0.9,1]$ or $[0,1]$ show significantly different results, which are excluded from the discussion here.

For evaluation, we generated 1080 sets of requests and network parameters such as switch node capacity and optical fiber capacity, and input them into each routing formulation for scheduling. Each scheduled routing path is evaluated with the same formula to calculate its fidelity: multiply each fidelity along the path, and apply the effect of error correction using the simplified functions as in~\cite{jouzdani2014fidelity}. For entanglement purification networks, the effect of purification is applied using the recursive functions as in~\cite{li2022fidelity}. For each set of requests, the result of average fidelity is calculated based on the fidelity of each scheduled (s,d) routing path weighted by its number of transferred qubits. The combined results of all requests for each model are shown in Fig.~\ref{fig:eval}.

As shown in Fig.~\ref{fig:eval}(a)~\ref{fig:eval}(b), the error correction cycles within servers significantly boost the overall fidelity of each communication, achieving a great performance increase compared to the baseline model Raw, in which no error correction is performed along the routing path. In Fig.~\ref{fig:eval}(c)~\ref{fig:eval}(d), we can see a more comprehensive comparison of all models. In general, our SurfaceNet model outperforms the entanglement purification model when its number of entanglement pairs for purification remains small (N$=$ 1, 2); meanwhile, although the entanglement purification model can employ more entanglement pairs (e.g., N$=$ 9) to increase its fidelity, it is highly impractical in real-world settings, as the entanglement generation is probabilistic and only has a short life span, which makes the large scaling on entanglement pairs very difficult. On the contrary, our SurfaceNet does not rely on entanglements between nodes, so it can be highly practical for real implementation. Also note that if we follow a similar approach as entanglement purification network that we perform an error correction at each node, the result fidelity of each path will trivially be 1, while similar to the N$=$ 9 case, this approach is impractical.

\begin{figure}[t]
\color{black}
\centerline{\includegraphics[scale=.56]{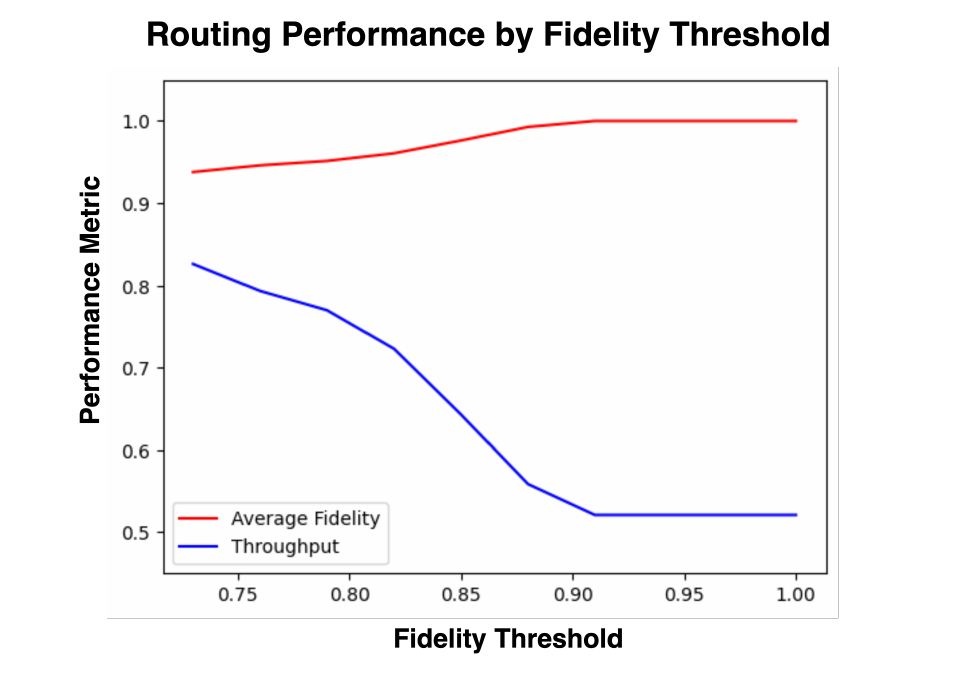}}
\caption{Average fidelity and throughput of communications in SurfaceNet with different fidelity threshold for individual communication. Both average fidelity and throughput range within $[0,1]$, for which higher is better.}
\label{fig:eval_c}
\end{figure}

Despite the analysis on reliability of the network, we also compared the throughput of each model, yet they differ slightly, which is reasonable since the relationship between fidelity and throughput can always be viewed as a trade-off, as one can always decrease the threshold of fidelity to allow for more pairs of comparatively unstable communication. 
\textcolor{black}{As shown in Fig.~\ref{fig:eval_c}, we changed the fidelity threshold for the fidelity constraint in SurfaceNet. Low fidelity threshold allows for low-fidelity communications, so its throughput is high while the average fidelity of each communication is low; higher fidelity threshold is more selective for communication quality, so its throughput becomes lower while its average fidelity becomes higher.}

\textcolor{black}{
During the simulation, we also observed that the location of server nodes plays a critical role in the overall performance of SurfaceNet, especially when it comes to ensuring the accuracy of communications since error corrections can only take place within the servers. 
In general, distributing server nodes evenly throughout the network layout results in enhanced performance, whereas loosely placing them in sparsely populated areas within the network leads to poor performance.
However, the precise mathematical relationship between server locations and network performance requires further investigation.
}
\\

\section{Related Work}

\textcolor{black}{
Quantum networks emerged as a granted solution to interconnect remote quantum devices. Current physical quantum networks~\cite{Wehner2018QI} remains in its early stages at small scales, limited by contemporary techniques of qubits processing and storage, also the high costs of establishing repeater nodes and photonic channels.
Much has been done in theoretical designs of quantum networks based on quantum teleportation and quantum entanglements. 
Many research groups have focused on the general architecture designs such as the repeaters or routers~\cite{munro2015inside} 
or the photonic channels~\cite{tengner2008photonic}.
}

\textcolor{black}{
Significant work has also been done in quantum network routing~\cite{briegel1998quantum, chen2022heuristic, zeng2022multi} equipped with techniques of entanglement purification~\cite{li2022fidelity} to improve the quality of entangled pairs shared across inferior or distant optical fibers. In the field of fault-tolerant communication~\cite{fowler2010surface, muralidharan2014ultrafast}, different quantum codes~\cite{gottesman1997stabilizer} have been proposed to serve as logical qubits. For surface code, different decoders~\cite{fowler2012surface} have been proposed to improve the error threshold of surface codes, and thus reduce the requirement for the fidelity at each optical fiber.
}
\\

\section{Conclusion}

In this paper, we presented SurfaceNet, a quantum network that employs surface codes as the logical qubits to preserve and transfer message. By utilizing the error correction capabilities of surface codes, our network ensures fault tolerance. We proposed a novel one-way communication procedure that allows for partitioning of individual surface codes and simultaneous transmission via multiple paths. The parts can then be merged back together to perform error corrections. Based on this new communication procedure, we also proposed a routing protocol that optimizes resource utilization of the network while maintaining a fidelity threshold for each communication, serving as a control to balance the efficiency and the reliability of our network. To evaluate the effectiveness of our routing protocol, we conducted simulations on randomly generated networks using an Integer Programming approximation for our routing protocol. The results showed that SurfaceNet significantly improves the overall fidelity of each communication.

\newpage
\bibliographystyle{ieeetr}
\bibliography{main} 

\end{document}